\documentstyle[12pt]{article}

\begin{document}

\baselineskip=18pt 

\begin{titlepage}

\begin{flushright}
UR-1383, ER-40685-832 \\
November 1994
\end{flushright}

\medskip

\begin{center}

{\bf PARITY VIOLATION IN TOP QUARK PAIR PRODUCTION \\
AT THE FERMILAB TEVATRON COLLIDER}

\vspace{0.36in}

CHUNG KAO\footnote{Internet Address: KAO@URHEP.PAS.ROCHESTER.EDU}

{\sl Department of Physics and Astronomy, University of Rochester \\
Rochester, NY 14627, USA}

\end{center}

\vspace{0.36in}

\begin{abstract}

The leading weak corrections to the production of top quark pairs
via $q\bar{q} \to t\bar{t}$ in $p\bar{p}$ collisions are evaluated.
The chromo-anapole form factor of the top quark
and effects of parity violation are studied
in the Standard Model (SM) and the Minimal Supersymmetric Model (MSSM).
The parity violation effect in $q\bar{q} \to g \to t\bar{t}$
from the SM weak corrections is found to be very small.
In the MSSM, the effects of parity violation
can be enhanced for $\tan\beta \equiv v_2/v_1 > \sqrt{m_t/m_b}$.

\end{abstract}

\end{titlepage}

\baselineskip=24pt 

\noindent{\bf 1. Introduction}

\medskip

In the Standard Model (SM) and some of its extensions,
the Yukawa couplings of a fermion to elementary spin-0 bosons
are proportional to the fermion mass.
The top quark now appears to be very heavy \cite{D0,CDF};
therefore, its Yukawa couplings are interestingly large.
Many processes involving the top quark and the `Higgs bosons'
might provide opportunities to investigate physics of
the electroweak symmetry breaking.

The QCD corrections \cite{Nason}-\cite{Laenen}
to the production rate of $q\bar{q} \to t\bar{t}$
are larger than the weak corrections \cite{Stange}-\cite{KLY2}.
However, the manifestation of parity violation in the electroweak interactions
might produce effects unobtainable from QCD,
which conserves both parity ($P$) and charge conjugation ($C$) symmetries.
In the SM, the dominant decay mode of the top quark is $t \to W^+ b$.
Since the top quark appears to be so heavy, its lifetime
is much shorter than the time needed to flip its spin \cite{Bigi}.
The helicity of the top quark can be deduced from the energy
distribution of the $W$'s or the leptons in the $W$ decays.
It was recently suggested \cite{KLY2} that an asymmetry
in the production rate of a right-handed top quark
associated with a left-handed top antiquark ($t_R\bar{t}_L$)
and $t_L\bar{t}_R$ can be a good observable of parity violation.
This asymmetry from the SM weak corrections was found to be small.

In this paper, the leading weak corrections to $q\bar{q} \to g \to t\bar{t}$
are evaluated in the Minimal Supersymmetric Model (MSSM)\footnote{
Reviews for the MSSM can be found in \cite{Nilles}-\cite{Tata}.
Recent studies on the search for MSSM Higgs bosons
are to be found in \cite{Barger}-\cite{Dai}. }
as well as in the SM.
The chromo-anapole form factor of the top quark is calculated
to study effects of parity violation in top quark pair production
at the Fermilab Tevatron, where $q \bar{q}$ annihilation
dominates the production of $t\bar{t}$.
Gluon fusion ($gg \to t\bar{t}$) will be the major source
of SM top quark pairs at the Large Hadron Collider (LHC)\footnote{
It might be possible that some large new physics effects
will appear in the process of $q\bar{q} \to t\bar{t}$ \cite{Hill}
and $gg \to t \bar{t}$ \cite{Eichten} at the LHC.}.
Parity violation in $t\bar{t}$ production via $q \bar{q}$ annihilation
and gluon fusion at the LHC is currently under investigation.
The parity conserving corrections from QCD and QED
do not affect the parity violating asymmetries; therefore,
they are not included in our analysis.

\bigskip

\noindent{\bf 2. The Top Quark Chromo-anapole Form Factor}

\medskip

In the SM, the $gt\bar{t}$ vertex gets
one-loop weak corrections from the Higgs boson ($H^0$),
the neutral and charged Nambu-Goldstone bosons ($G^0$, $G^\pm$),
the $Z$ boson ($Z$) and the $W$ bosons ($W^\pm$).
The Feynman diagrams are shown in Figure 1.
In the 't~Hooft-Feynman gauge, the weak interaction Lagrange density
involving the top quark is
\begin{eqnarray}
{\cal L}_I
& = &-\bar{t}\gamma^\mu(g_V-g_A\gamma_5) t Z_\mu  \nonumber \\
&   &-\frac{g}{2\sqrt{2}}
       [ \bar{t}\gamma^\mu(1-\gamma_5) b W^+_\mu
        +\bar{b}\gamma^\mu(1-\gamma_5) t W^-_\mu ] \nonumber \\
&   &-\frac{m_t}{v}\bar{t}t H^0 +i\frac{m_t}{v}\bar{t}\gamma_5 t G^0
      \nonumber \\
&   &+\frac{m_t}{\sqrt{2} v}
     [\bar{t} (\lambda^G_s-\lambda^G_p\gamma_5) b G^+
     +\bar{b} (\lambda^G_s+\lambda^G_p\gamma_5) t G^- ]
\end{eqnarray}
where
$g_V = \frac{g}{4 \cos\theta_W}[1 -\frac{8 \sin^2\theta_W}{3}] \sim 0.08$,
$g_A = \frac{g}{4 \cos\theta_W} \sim 0.19$,
$\lambda^G_s = 1 -\frac{m_b}{m_t}$,
$\lambda^G_p = 1 +\frac{m_b}{m_t}$,
$v$ is the vacuum expectation value (VEV) of the Higgs doublet
and $\theta_W$ is the Weinberg weak mixing angle.
We take $V_{tb} = 1$ for simplicity.
The quark mixing from the CKM matrix is not included; thus, CP is conserved.

In the Minimal Supersymmetric Model (MSSM), there are two Higgs doublets
and five Higgs bosons: a pair of singly charged Higgs bosons, $H^{\pm}$;
two neutral CP-even Higgs bosons, $H$ (heavier) and $h$ (lighter);
and a CP-odd Higgs boson, $A$.
The Lagrange density involving the top quark and the Higgs bosons
in the MSSM is
\begin{eqnarray}
{\cal L}_I
& = &-[\frac{\sin\alpha}{\sin\beta}] \frac{m_t}{v} \bar{t}t H
     -[\frac{\cos\alpha}{\sin\beta}] \frac{m_t}{v} \bar{t}t h \nonumber \\
&   &+i [\cot\beta] \frac{m_t}{v} \bar{t}\gamma_5 t A \nonumber \\
&   &+\frac{m_t}{\sqrt{2} v}
      [\bar{t}(\lambda^H_s-\lambda^H_p\gamma_5) b H^+
      +\bar{b}(\lambda^H_s+\lambda^H_p\gamma_5) t H^- ]
\end{eqnarray}
where $\tan\beta \equiv v_2/v_1$ is the ratio of the two Higgs doublet VEV's,
$\lambda^H_s = \cot\beta +\frac{m_b}{m_t} \tan \beta$,
$\lambda^H_p = \cot\beta -\frac{m_b}{m_t} \tan \beta$,
and $\alpha$ is the mixing angle between the neutral CP-even Higgs bosons.
At the tree level, all masses and couplings of Higgs bosons
can be determined with only two independent parameters,
which we choose to be $\tan \beta$
and the mass of the charged Higgs boson $M_{H^+}$.
We have included one loop corrections from top and bottom
Yukawa interactions to the Higgs masses and couplings
using the effective potential \cite{OYY}-\cite{Haber}.
For simplicity, we assume that all squarks have the same mass $m_{Q~} = 1$ TeV;
all neutralinos and charginos are very heavy;
and there is no mixing between the top squarks $\tilde{t}_L$ and $\tilde{t}_R$.

The 't~Hooft-Feynman gauge is employed to evaluate the loop diagrams
with $M_{G^0} = M_Z$ and $M_{G^+} = M_W$.
We take $M_W = 80.22$ GeV, $M_Z = 91.187$ GeV,
$m_b = 4.8$ GeV, $m_t = 170$ GeV and $\sin^2\theta_W = 0.2319$.
The updated parton distribution functions of CTEQ3L \cite{CTEQ}
are chosen to evaluate the cross section of $p\bar{p} \to t\bar{t} +X$
with $\Lambda = 0.177$ GeV and $Q^2 = \hat{s}$.

Let us write the $gt\bar{t}$ vertex as
\begin{equation}
-ig_s \bar{u}(p) T^a \Gamma^\mu v(q)
\end{equation}
where $g_s$ is the strong coupling, $T^a$ are the $SU_C (3)$ matrices,
$u(p)$ and $v(q)$ are the Dirac spinors of $t$ and $\bar{t}$
with outgoing momenta $p$ and $q$.
At the tree level, $\Gamma_0^\mu = \gamma^\mu$.
The 1-loop vertex function can be expressed
in terms of form factors \cite{KLY0}
\begin{eqnarray}
\Gamma_1^{\mu} & = & \gamma^\mu[A(k^2)-B(k^2)\gamma_5] \nonumber \\
               &  +& (p-q)^\mu [C(k^2)-D(k^2)\gamma_5] \nonumber \\
               &  +& (p+q)^\mu [E(k^2)-F(k^2)\gamma_5]
\end{eqnarray}
where $k = p+q$,
$k^2 = M_{t\bar{t}}^2 = \hat{s}$,
and $M_{t\bar{t}}$ is the invariant mass of $t\bar{t}$.
The conservation of vector current demands that
$B(k^2) = -k^2 F(k^2)/(2m_t)$ and $E(k^2) = 0$.
Applying the Gordon identities, we obtain
\begin{eqnarray}
\Gamma_1^{\mu}
& = & F_1(k^2) \gamma_\mu -F_2(k^2) i\sigma^{\mu\nu} k_\nu \nonumber \\
&   & +a(k^2)  \gamma_\nu \gamma_5 ( k^2 g^{\mu\nu} -k^\mu k^\nu)
      +d(k^2)  i\sigma^{\mu\nu} k_\nu \gamma_5
\end{eqnarray}
Comparing both expressions for the $\Gamma_1^{\mu}$, we obtain
(1) $F_1(k^2) = A(k^2) +2 m_t C(k^2)$, $F_1(0) =$ the chromo-charge;
(2) $F_2(k^2) = C(k^2)$, $F_2(0) =$ the anomalous chromo-magnetic moment;
(3) $a(k^2) = -B(k^2)/k^2 = F(k^2)/2m_t$, $a(0) =$ chromo-anapole moment;
and (4) $d(k^2) = D(k^2)$, $d(0) =$ the chromo-electric dipole moment.
The quark mixing from the CKM matrix is not included;
therefore, the electric dipole form factor $D(k^2)$ is equal to zero
in our analysis.

A nonzero anapole form factor generates signatures of parity violation.
In the SM, the chromo-anapole form factor $B_{SM}$ has contributions from
the $Z$ ($B_Z$), the $W^+$ ($B_W$), and the $G^+$ ($B_G$).
Only the real parts of the form factors appear in the observables
that we will discuss.
Therefore, we present the real part of the form factor $B(k^2)$
in Figure 2 for every diagram and the SM total.
Since the form factor $B(k^2)$ is proportional to $\hat{s} = M_{t\bar{t}}^2$,
the parity violation effect will be greatly enhanced at higher energy.
The $Z$ boson contributes very little to the anapole form factor of
the top quark since $2g_V g_A$ is small.
In the 't~Hooft-Feynman gauge, $|Re(B_G)|$ is larger than the full
$|Re(B_{SM})|$, because the $B_G$ and the $B_W$ have opposite signs.
The total $B_{SM}$ appears to be small.

In the MSSM, the chromo-anapole form factor $B_{MSSM}$
gets an additional contribution ($B_{H}$) from the charged Higgs boson.
The form factor $B_{H}$ is a function of $M_{H^+}$ and $\tan \beta$.
It is proportional to $\lambda^H = \lambda_s^H \lambda_p^H
= \cot^2\beta\{1 -[(m_b/m_t)\tan^2\beta]^2\}$.
For $M_{t\bar{t}}$ larger than about 400 GeV,
if $\tan\beta > \sqrt{m_t/m_b}$, then $\lambda^H <0$
and the $B_H$ and the $B_{SM}$ interfere constructively.
On the other hand, if $\tan\beta < \sqrt{m_t/m_b}$,
then the $B_H$ and the $B_{SM}$ interfere destructively.
When $M_{t\bar{t}}$ is close to the $2m_t$ threshold,
the relative sign between the $B_H$ and the $B_{SM}$
is opposite to the relative sign for $M_{t\bar{t}}$ above 400 GeV.
For $\tan\beta = \sqrt{m_t/m_b}$, $B_H$ vanishes
and the $B_{MSSM}$ is equal to the $B_{SM}$.
Figure 3 shows the form factor $B_{MSSM}$
for various values of $M_{H^+}$ and $\tan\beta$.

\bigskip

\noindent{\bf 3. Asymmetries and Parity Violation}

\medskip

Let us define the cross section of the subprocess $q\bar{q} \to t\bar{t}$
in each helicity state of the $t\bar{t}$ as
\begin{equation}
\hat{\sigma}_{\lambda_1,\lambda_2}
\equiv \hat{\sigma}( q\bar{q} \to t_{\lambda_1} \bar{t}_{\lambda_2} )
\end{equation}
where $\lambda_{1,2}$ represents a right-handed ($R$)
or a left-handed ($L$) helicity.
At the tree level,
\begin{eqnarray}
\hat{\sigma}^{(0)}_{LL} & = & \hat{\sigma}^{(0)}_{RR}
 =   \frac{ 4 \pi \alpha_s^2 \beta}{27\hat{s}^2} (2m_t^2) \nonumber \\
\hat{\sigma}^{(0)}_{LR} & = & \hat{\sigma}^{(0)}_{RL}
 =   \frac{ 4 \pi \alpha_s^2 \beta}{27\hat{s}^2} (\hat{s}) \nonumber \\
\hat{\sigma}^{(0)}
 & = & \hat{\sigma}^{(0)}_{LL} +\hat{\sigma}^{(0)}_{RR}
      +\hat{\sigma}^{(0)}_{LR} +\hat{\sigma}^{(0)}_{RL}
   =  \frac{ 8 \pi \alpha_s^2 \beta}{27\hat{s} }( 1+2m_t/\hat{s} )
\end{eqnarray}
where the cross section has been summed and averaged over spins and colors
of the quarks in the initial state.

With one loop weak corrections, the cross section
in each helicity state of $t\bar{t}$ is
$\hat{\sigma}_{\lambda_1,\lambda_2} =  \hat{\sigma}^{(0)}_{\lambda_1,\lambda_2}
+ \delta \hat{\sigma}_{\lambda_1,\lambda_2}$;
where $\delta \hat{\sigma}_{\lambda_1,\lambda_2}$
is the contribution from weak corrections.
The $K$-factor in each helicity state of the $t\bar{t}$ is defined as
\begin{eqnarray}
K_{\lambda_1,\lambda_2}
& \equiv & \frac{\hat{\sigma}_{\lambda_1,\lambda_2}}
                {\hat{\sigma}^{(0)}_{\lambda_1,\lambda_2}}
        =  1 +\frac{\delta \hat{\sigma}_{\lambda_1,\lambda_2}}
                   {\hat{\sigma}^{(0)}_{\lambda_1,\lambda_2}}
\end{eqnarray}
These $K$-factors can be expressed in terms of the form factors:
\begin{eqnarray}
K_{LL}& = &1 +2Re(A) -\beta^2 \hat{s} Re(C)/m_t +\beta \hat{s} Re(D)/m_t
           \nonumber \\
K_{RR}& = &1 +2Re(A) -\beta^2 \hat{s} Re(C)/m_t -\beta \hat{s} Re(D)/m_t
           \nonumber \\
K_{LR}& = &1 +2Re(A) +2\beta Re(B)
           \nonumber \\
K_{RL}& = &1 +2Re(A) -2\beta Re(B)
\end{eqnarray}
where $\beta = \sqrt{ 1 -4m_t^2/\hat{s} }$.
Figure 4 shows the $K$ factors in the MSSM with $M_{H^+} =$ 180 GeV
for $\tan\beta =$ 1 and 35, and in the SM with $M_{H^0} =$ 100 GeV.
The quark mixing from the CKM matrix is not included;
therefore, CP is invariant and $K_{LL}$ is equal to $K_{RR}$.
The difference in $K_{RL}$ and $K_{LR}$ is generated by
parity violating corrections.

The cross section of $p\bar{p} \to t\bar{t} +X$
is evaluated with the convolution
of the subprocess cross section ($\hat{\sigma}(q\bar{q} \to t\bar{t})$)
and parton distribution functions.
The effects of parity violation appear as an asymmetry
in the invariant mass distributions as well as
in the integrated cross sections of $t_R\bar{t}_L$ and $t_L\bar{t}_R$.
Let us define the differential asymmetry as
\begin{eqnarray}
\delta {\cal A}(M_{t\bar{t}})
& \equiv &\frac{d\sigma_{RL}/dM_{t\bar{t}} -d\sigma_{LR}/dM_{t\bar{t}}}
          {d\sigma_{RL}/dM_{t\bar{t}} +d\sigma_{LR}/dM_{t\bar{t}}} \nonumber \\
& = &\frac{ K_{RL}-K_{LR} }{ K_{RL}+K_{LR} } \nonumber \\
& = &\frac{-2\beta Re(B) }{ 1 +2Re(A) }
\end{eqnarray}
where the parton distribution functions are cancelled in the ratio.
This differential asymmetry is presented in Figure 5 for the SM and the MSSM.
The form factor $B(k^2)$ is proportional to $\hat{s} = M_{t\bar{t}}^2$;
therefore, the differential asymmetry is enhanced at higher energy.

Since the helicity of the top quark has to be deduced from the energy
distribution of the leptons in the $W$ decays,
it is more realistic to sum over the helicities of the $\bar{t}$
and consider an integrated asymmetry in numbers of $t_R$ and $t_L$.
The number of observed $t\bar{t}$ will be reduced only by
by the branching ratio of $t \to W^+ b \to l^+\nu_l b$.
The integrated asymmetry is defined as
\begin{eqnarray}
{\cal A}
  & \equiv &\frac{ N_{R} -N_{L} }{ N_{R} +N_{L} }
     = \frac{ \sigma_{R} -\sigma_{L} }{ \sigma_{R} +\sigma_{L} } \nonumber \\
N & = & \cal{L} \sigma
\end{eqnarray}
where $\sigma_{R} = \sigma_{RL} +\sigma_{RR}$,
$\sigma_{L} = \sigma_{LR} +\sigma_{LL}$
and $\cal{L} =$ the integrated luminosity.
The statistical uncertainty ($\Delta{\cal A}$)
and statistical significance ($N_S$) of the integrated asymmetry are
\begin{eqnarray}
\Delta{\cal A} & =    & \frac{ 2\sqrt{N_{R}N_{L}} }{ (N_{R}+N_{L})^{3/2} }
                \sim  \frac{1}{ \sqrt{N_{R}+N_{L}} } \nonumber \\
N_S & = & {\cal A}/\Delta{\cal A}.
\end{eqnarray}
To be conservative, we require $N_S \ge 4$ for an asymmetry to be
possibly visible.

The difference ($\Delta \sigma$) and the total ($\sigma$)
of the cross sections $\sigma_R$ and $\sigma_L$
with one loop SM weak corrections,
as well as the integrated asymmetry (${\cal A}$)
and its statistical significance ($N_S$) are presented in Table I
for an integrated luminosity of 10 fb$^{-1}$ and 100 fb$^{-1}$.
At $\sqrt{s} =$ 2 TeV, the asymmetry will not be visible
even if $\cal{L} =$ 100 fb$^{-1}$.
At $\sqrt{s} =$ 4 TeV, the asymmetry can become larger than its uncertainty
with $\cal{L} =$ 100 fb$^{-1}$.
We find that requiring $M_{t\bar{t}} >$ 500 GeV
can efficiently enhance the asymmetry
while only slightly increasing its uncertainty.

In the MSSM, the asymmetry can be enhanced if $\tan \beta > \sqrt{m_t/m_b}$.
Table II shows the integrated asymmetry
and its statistical significance in the MSSM for $M_{t\bar{t}} >$ 500 GeV.
If $\tan \beta$ is close to one, the asymmetry ${\cal A}$ from the MSSM
weak corrections is smaller than that in the SM.
For $10 > \tan \beta > 3$, the asymmetry generated from the MSSM
weak corrections is very similar to that in the SM.
At $\sqrt{s} =$ 2 TeV, the asymmetry $\cal{A}$ can be larger than
$\Delta \cal{A}$ for $\tan \beta > \sqrt{m_t/m_b}$
and $\cal{L} =$ 100 fb$^{-1}$.
At $\sqrt{s} =$ 4 TeV, the parity violation signal is highly enhanced.
With $\cal{L} =$ 100 fb$^{-1}$, this asymmetry might be visible
for $M_{H^+} \le$ 300 GeV and $\tan \beta \ge m_t/m_b \sim 35$.
Its statistical significance ($N_S$) can be larger than 4.8.

\bigskip

\noindent{\bf 4. Conclusions}

\medskip

In the SM, the asymmetry of parity violation
is almost independent of the Higgs boson mass.
At $\sqrt{s} = 2$ TeV, the asymmetry in $N_{R}$ and $N_{L}$ of $t\bar{t}$
is always smaller than its statistical uncertainty for
an integrated luminosity of 10 fb$^{-1}$.
It becomes slightly better with ${\cal L} =$ 100 fb$^{-1}$.
At $\sqrt{s} = 4$ TeV, the asymmetry is enhanced.
However, it is still difficult to observe this asymmetry
in the SM even with ${\cal L} =$ 100 fb$^{-1}$.

In the MSSM, the parity violation asymmetry depends mainly
on $\tan\beta$ and $M_{H^+}$.
It can be enhanced for $\tan\beta > \sqrt{m_t/m_b}$.
The asymmetry in $N_{R}$ and $N_{L}$ of $t\bar{t}$ might be visible
at $\sqrt{s} = 2$ and ${\cal L} =$ 100 fb$^{-1}$
or at $\sqrt{s} = 4$ TeV and ${\cal L} = $ 10 fb$^{-1}$,
if $\tan\beta \gg m_t/m_b$ and $M_{H^+}$ close to $m_t$.
At $\sqrt{s} = 4$ TeV, with ${\cal L} = $ 100 fb$^{-1}$,
this asymmetry generated from the MSSM weak corrections
might be visible for $M_{H^+} <$ 300 GeV and $\tan\beta$ close to $m_t/m_b$.

The parity violation in top quark pair production
might provide a good opportunity to study the parity violating interactions
between the top quark and spin-0 or spin-1 particles.
Because the parity violation signal is very small in the SM,
any observation of large parity violation would indicate new physics.

\bigskip

\noindent{\bf Acknowledgements}

\medskip

I am grateful to Duane Dicus, Lynne Orr, Xerxes Tata, especially,
Glenn Ladinsky and Chien-Peng Yuan for beneficial discussions,
comments and instructions.
This research was supported in part
by the US Department of Energy grant DE-FG02-91ER40685.


\newpage
%

\newpage

TABLE I.
The difference ($\Delta \sigma$) and the total ($\sigma$) of cross sections
$\sigma_R$ and $\sigma_L$,
and the asymmetry ${\cal A} = (N_{R}-N_{L})/(N_{R}+N_{L})$
generated by the SM weak corrections in $p\bar{p} \to t\bar{t} +X$,
with $m_t = 170$ GeV, for (a) $\sqrt{s} = 2$ TeV and (b) $\sqrt{s} = 4$ TeV.
At each energy, we consider both $M_{t\bar{t}} > 2m_t$,
and $M_{t\bar{t}} > $ 500 GeV.
Also shown is the statistical significance ($N_S$)
for ${\cal L} =$ 10 fb$^{-1}$ and 100 fb$^{-1}$.

\medskip

\begin{center}
\begin{tabular}{ccccccc}
\hline
$M_H$ & $\Delta\sigma$ & $\sigma$ & ${\cal A}$  & $N_S$ & $N_S$ \\
(GeV) & (fb) & (fb) & ($\%$)  &(10 fb$^{-1})$ &(100 fb$^{-1})$ \\
\hline
(a) $\sqrt{s}$ = 2 TeV \\
$M_{t\bar{t}} > 2m_t$ \\
100 & 3.76 & 4550 & 0.083 & 0.18 & 0.56 \\
700 & 3.76 & 4510 & 0.083 & 0.18 & 0.56 \\
$M_{t\bar{t}} > 500$ GeV \\
100 & 3.00 &  845 & 0.36  & 0.33 & 1.0 \\
700 & 3.00 &  858 & 0.35  & 0.32 & 1.0 \\
(b) $\sqrt{s}$ = 4 TeV \\
$M_{t\bar{t}} > 2m_t$ \\
100 & 25.6 & 16240 & 0.16 & 0.63 & 2.0 \\
700 & 25.6 & 16170 & 0.16 & 0.64 & 2.0 \\
$M_{t\bar{t}} > 500$ GeV \\
100 & 22.5 & 4860 & 0.46 & 1.0 & 3.2 \\
700 & 22.5 & 4940 & 0.46 & 1.0 & 3.2 \\
\hline
\end{tabular}
\end{center}
%

\newpage

TABLE II.
The difference and the total of cross sections $\sigma_R$ and $\sigma_L$,
and the asymmetry ${\cal A}$ generated by the MSSM weak corrections,
for $M_{t\bar{t}} > $ 500 GeV
as well as several values of $M_{H^+}$ and $\tan \beta$.
All notations and other parameters are the same as in Table I.
\medskip

\begin{center}
\begin{tabular}{cccccc}
\hline
$\tan\beta$ & $\Delta\sigma$ & $\sigma$ & ${\cal A}$  & $N_S$ & $N_S$ \\
            & (fb) & (fb) & ($\%$)  &(10 fb$^{-1})$ &(100 fb$^{-1})$ \\
\hline
(a) $\sqrt{s} = 2$ TeV \\
$M_{H^+} =$ 180 GeV \\
1  & -1.11 & 835 & -0.13 & 0.12 & 0.38 \\
3  &  2.57 & 846 &  0.30 & 0.28 & 0.89 \\
10 &  3.29 & 847 &  0.39 & 0.36 & 1.1 \\
35 &  7.02 & 839 &  0.84 & 0.77 & 2.4 \\
$M_{H^+} =$ 300 GeV \\
1  & 1.37 & 844 & 0.16 & 0.15 & 0.47 \\
3  & 2.83 & 847 & 0.33 & 0.31 & 0.97 \\
10 & 3.12 & 848 & 0.37 & 0.34 & 1.1 \\
35 & 4.60 & 847 & 0.54 & 0.50 & 1.6 \\
(b) $\sqrt{s} = 4$ TeV \\
$M_{H^+} =$ 180 GeV \\
1  & -4.08 & 4790 & -0.085 & 0.19 & 0.59 \\
3  &  19.7 & 4860 &  0.41 & 0.90 & 2.8 \\
10 &  24.4 & 4870 &  0.50 & 1.1  & 3.5 \\
35 &  48.5 & 4820 &  1.0  & 2.2  & 7.0 \\
$M_{H^+} =$ 300 GeV \\
1  & 10.9 & 4840 & 0.23 & 0.50 & 1.6 \\
3  & 21.3 & 4870 & 0.44 & 0.97 & 3.1 \\
10 & 23.3 & 4880 & 0.48 & 1.1  & 3.3 \\
35 & 33.8 & 4860 & 0.70 & 1.5  & 4.8 \\
$M_{H^+} =$ 500 GeV \\
1  & 20.8 & 4860 & 0.43 & 0.94 & 3.0 \\
3  & 22.3 & 4870 & 0.46 & 1.0  & 3.2 \\
10 & 22.6 & 4880 & 0.46 & 1.0  & 3.2 \\
35 & 24.2 & 4880 & 0.50 & 1.1  & 3.5 \\
\hline
\end{tabular}
\end{center}
%

%
\newpage
\noindent{\bf Figures}

\bigskip

FIG. 1 The Feynman diagrams for weak corrections to
(a) the top quark self energy and wave function renormalization
and (b) the $gt\bar{t}$ vertex, in the SM and the MSSM.
Both momenta $p$ and $q$ are flowing out of the vertex.

\medskip

FIG. 2 The real part of the form factor $B(k^2)$ from diagrams
with the $Z$, the $G^+$, the $W^+$, and the SM total,
as a function of $M_{t\bar{t}}$ for $m_t =$ 170 GeV.

\medskip

FIG. 3 The real part of the form factor $B_{MSSM}$
as a function of $M_{t\bar{t}}$, for (a) $M_{H^+} =$ 180 GeV
and (b) $M_{H^+} =$ 300 GeV, with $\tan\beta =$ 1, 3, 10, and 35.
Also shown is real part of the form factor $B_{SM}$  which is
independent of the Higgs boson mass.

\medskip

FIG. 4 The $K$-factor as a function of $M_{t\bar{t}}$,
at various helicity states of $t\bar{t}$,
(a) in the SM with $M_{H^0} =$ 100 GeV,
as well as in the MSSM with $M_{H^+} =$ 180 GeV for
(b) $\tan\beta =$ 1, and (c) $\tan\beta =$ 35.
The first letter indicates the helicity of the the $t$,
the second letter indicates that of the $\bar{t}$.

\medskip

FIG. 5 The asymmetry in the invariant mass distribution of $t\bar{t}$
as defined in Eq. 10,
for (a) $M_{H^+} =$ 180 GeV and (b) $M_{H^+} =$ 300 GeV,
with $\tan\beta =$ 1, 3, 10, and 35.
Also shown is the same asymmetry in the SM with $M_{H^0} =$ 100 GeV (solid)
and $M_{H^0} =$ 500 GeV (dash).

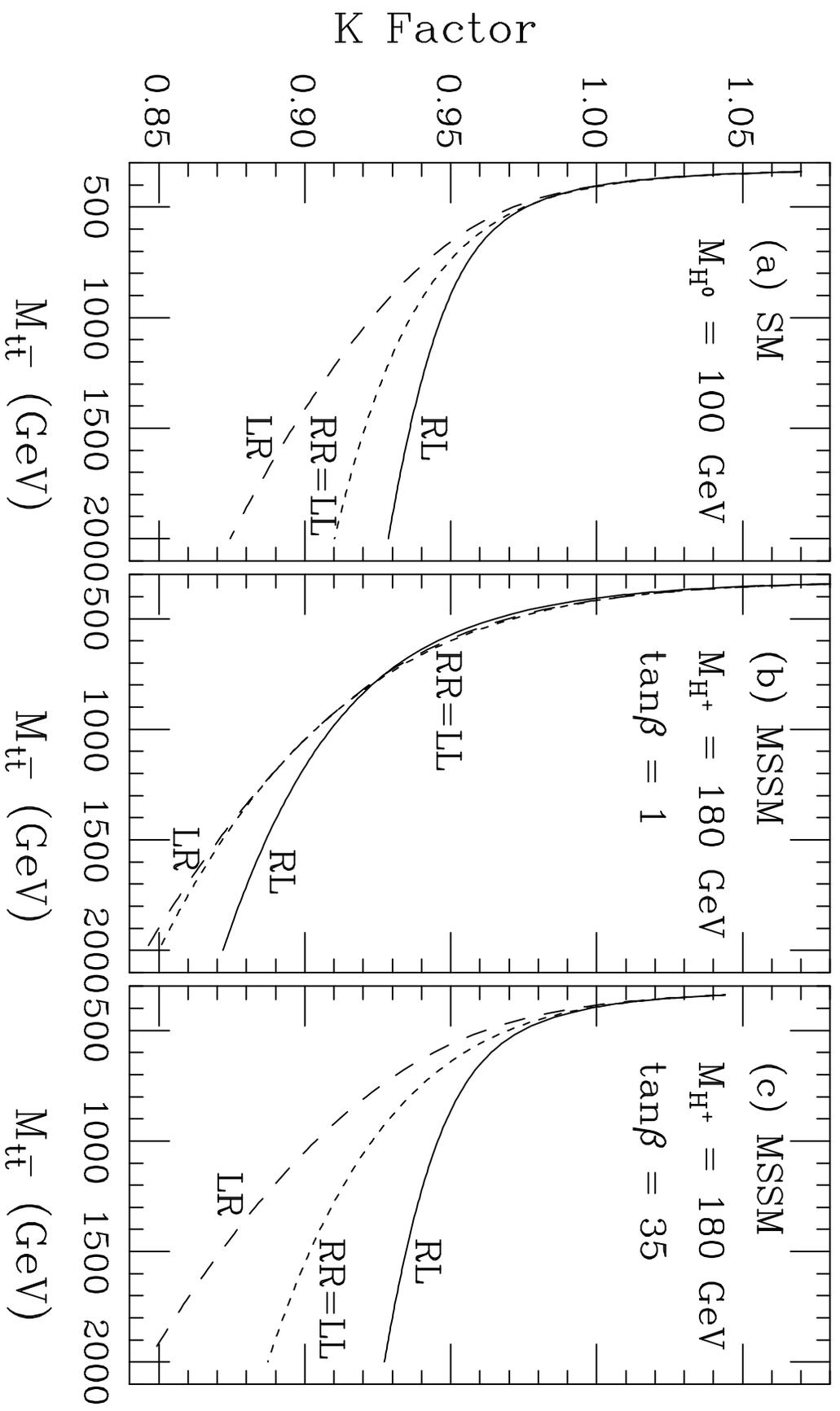
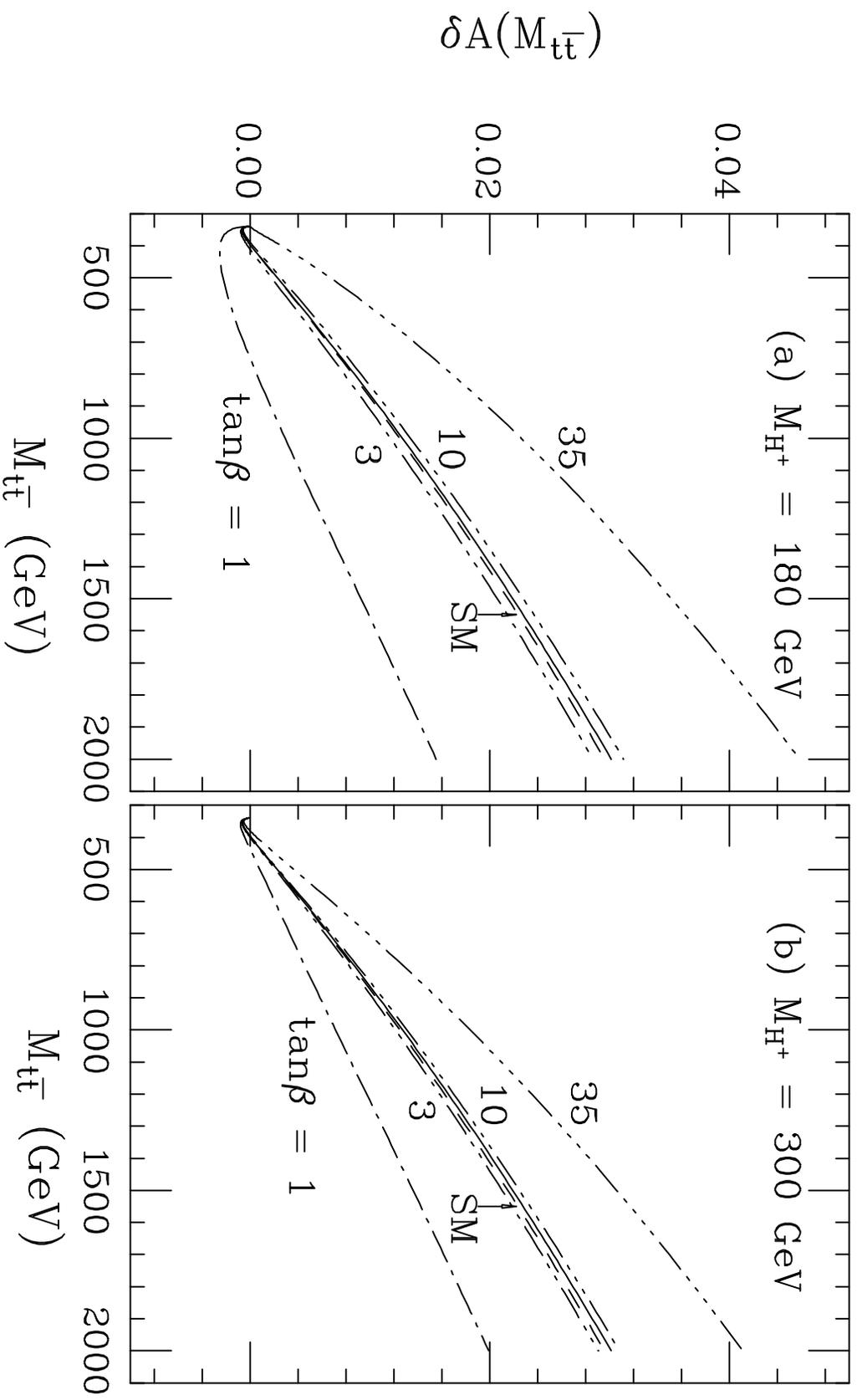
\end{document}